# Machine learning optimization of the collocation point set for solving the Kohn-Sham equation

Jonas Ku[a], Aditya Kamath[a], Tucker Carrington Jr[b], Sergei Manzhos[c,1]

[a] Department of Mechanical Engineering, National University of Singapore, Block EA #07-08, 9 Engineering Drive 1, Singapore 117576 Singapore.
[b] Chemistry Department, Queen's University, Kingston, Ontario K7L 3N6, Canada
[c] Centre Énergie Matériaux Télécommunications, Institut National de la Recherche Scientifique, 1650, boulevard Lionel-Boulet, Varennes QC J3X1S2 Canada

**Abstract**

The rectangular collocation approach makes it possible to solve the Schrödinger equation with basis functions that do not have amplitude in all regions in which wavefunctions have significant amplitude. Collocation points can be restricted to a small region of space. As no integrals are computed, there are no problems due to discontinuities in the potential, and there is no need to use integrable basis functions. In this paper, we show, for the Kohn-Sham equation, that machine learning can be used to drastically reduce the size of the collocation point set. This is demonstrated by solving the Kohn-Sham equations for CO and $H_2O$. We solve the Kohn-Sham equation on a given effective potential which is a critical part of all DFT calculations, and monitor orbital energies and orbital shapes. We use a combination of Gaussian process regression and a genetic algorithm to reduce the collocation point set size by more than an order of magnitude (from about 51,000 points to 2,000 points) while retaining mHartree accuracy.

## 1 Introduction

In this paper, we propose new ideas for choosing the points used with rectangular collocation to solve the electronic Schrödinger equation using the Kohn-Sham equation as a test case. Rectangular collocation has been used to solve both the vibrational and the electronic Schrödinger equations.[1-13] We solve the time-independent Schrödinger equation (SE)

---

[1] Author to whom correspondence should be addressed. E-mail: sergei.manzhos@emt.inrs.ca ; Tel: + 1 514 2286900.

$$\widehat{H}\psi(x) \equiv (\widehat{T} + \widehat{V})\psi(x) = E\psi(x), \qquad (1)$$

where $\widehat{H}$ is the Hamiltonian operator, which is the sum of a kinetic energy operator (KEO) $\widehat{T}$ and the potential energy operator $\widehat{V}$; $\psi(x)$ is a wavefunction; and $E$ is an energy. It needs to be solved when computing electronic energy levels, as well as when computing (ro)vibrational spectra.[14-16] In the Born-Oppenheimer approximation assumed here, one can separately solve the electronic and nuclear Schrödinger equations. In the electronic case, $x$ is a vector of all the electronic coordinates (of which there are $3N_e$, where $N_e$ is the number of electrons considered in the model) and the potential term $\widehat{V} = V(x)$ is due to Coulombic interactions among electrons and nuclei. Considering $\psi(x)$ in Eq. 1 as an explicit function of the $3N_e$ (Cartesian) coordinates gives rise to various wavefunction-based *ab initio* approaches.[17-21] In this paper we solve a specific electronic Schrödinger equation: the single-electron Kohn Sham equation which is the basis of Kohn-Sham Density Functional Theory (DFT).[22,23] In the KS case, Eq. 1 takes the form,

$$-\frac{1}{2}\Delta\psi(x) + V_{eff}(x)\psi(x) = E\psi(x), \qquad (2)$$

where

$$V_{eff}(x) = V_{nuc}(x) + \int \frac{\rho(x')}{|x-x'|}dx' + V_{XC}(x)$$

$$\rho(x) = \sum_{i=1}^{N_e}|\psi_i(x)|^2, \quad \int_\infty |\psi_i(x)|^2 dx = 1, \qquad (3)$$

where $\rho(x)$ is the electron density, $V_{eff}(x)$ is an effective potential, $V_{XC}(x)$ is the exchange-correlation potential, $V_{nuc}(x)$ is the Coulombic potential due to the nuclei, $\psi_i(x)$ are the orbitals, and *i* indexes different solutions of Eq. 2. Here, $x$ is a vector of three Cartesian coordinates. We use atomic units (*a.u.*). Because Kohn-Sham DFT is computationally much less costly than wavefunction-based methods, it dominates *ab initio* materials modeling and much of molecular modeling. It is Eq. 2 that is solved in this article. We note that both $V(x)$ in wavefunction-based approaches and $V_{eff}(x)$ in DFT are singular due to the singular nature of the (Coulombic) ionic potential. Pseudopotentials can be used to remove the singularity at the cost of accuracy.[24]

The time-independent Schrödinger equation one must solve to compute a ro-vibrational spectrum of an isolated molecule has the form of Eq. 1, but the potential is obtained from the Born-Oppenheimer approximation, and the KEO, written in terms of coordinates that specify the shape and orientation of the molecule, is complicated. Cartesian coordinates can be used when translation and rotation are frustrated.

As discussed in section 2, collocation is useful for solving the electronic Schrödinger equation because it obviates the need to choose quadrature points with which integrals can be approximated. In particular, because there are no integrals, singularities and slow potential decay in the asymptotic region pose no problems. Methods related to collocation have been previously used to solve the electronic Schrödinger equation. In some cases,[12,13,25-27] analytic integrals are used for some matrix elements; this is fundamentally different from our approach in which there are no integrals and all matrix elements are determined by evaluating functions at points. In other cases,[12,28-30] collocation is used, but with complicated basis functions.

In a recent work,[31] we have shown that rectangular collocation can be effective for solving the electronic Schrödinger equation. We explicitly excluded points from the cusp region and concentrated collocation points around atoms without sampling the exponential tails of the wavefunction. We showed that this point placement strategy is advantageous compared to a uniform grid. Slater-type basis functions were used, which is difficult with a variational method. We solved the Schrödinger equation for the H atom and $H_2^+$ and the Kohn-Sham equation for CO and $H_2O$. The same ideas will also work when the operator on the left in Eq. 2 is the Fock operator. When solving the KS equation, to achieve our target mHa accuracy, we used about $M = 51,000$ collocation points.[31] A linear combination of atomic orbitals (LCAO) type basis was used with about $N = 150$ basis functions, i.e. we had $M \gg N$. Optimization of collocation points and of the basis can greatly improve accuracy or decrease the calculation cost, as we have shown for the case of the vibrational Schrödinger equation.[3-7,9,10] In Ref. 31, the point set was not optimized.

*In this paper, we optimize the collocation point set to decrease the cost of solving the Kohn-Sham equation.* We solve Kohn-Sham equation on a given effective potential – a critical step in any DFT calculation. This step produces orbitals and their energies. In a complete DFT calculation, to achieve self-consistency, the orbitals and energies are used to assemble a new effective potential and density for the next step and for total energy calculations. This is done with well-established procedures and is not considered here. Orbital energies themselves are also explicitly used in many types of simulations including linear-response time-dependent DFT calculations,[32] estimations of ionization potentials (IP) and electron affinities (EA) including higher IP and lower EA;[33] in applications, they are also widely used to approximate redox potentials and driving forces to electron transfer.[34,35] In addition, electronic coupling is often estimated based on orbital energy splitting[36] and energies of core electronic orbitals are helpful in the analysis of XPS spectra.[37] We therefore focus on orbital energies in order to optimize collocation points. We optimize the collocation points by choosing them to minimize the root means square error (*rmse*) of the potential obtained by representing the exact potential with Gaussian processes (GP).[38,39] We hypothesize that points that are the best for representing the potential are also good for solving the SE. This is strikingly demonstrated by calculations on formaldehyde, where it was shown that it is possible to compute accurate vibrational levels using only points on the slabs required to determine a multimode representation[40-43] of the PES.[5] It was also shown in Ref. 11 where the fitting points (used with GP and neural network PES representations) that reduced the global PES error also reduced the error in vibrational levels computed with collocation. To minimize the potential *rmse*, we use a genetic algorithm (GA).[44] A GA is used to avoid the problem of local minima. We feed the GA populations obtained by optimizing the parameters of a probability distribution used to select them. We show, for the solution of the KS equation for CO and $H_2O$, that this scheme is capable of reducing the collocation point set size from about 51,000 points to only 2,000-3,000 points while maintaining mHa accuracy.

## 2   Collocation and its advantages

Both the electronic and the nuclear Schrödinger equations are typically solved by expanding the wavefunction in a basis:

$$\psi(x) \approx \sum_{k=1}^{N} c_k \phi_k(x), \qquad (4)$$

with a vector of coefficients $c$ and basis functions $\phi_k(x)$. The SE then becomes

$$\sum_{k=1}^{N} c_k \hat{T} \phi_k(x) + V(x) \sum_{k=1}^{N} c_k \phi_k(x) = E \sum_{k=1}^{N} c_k \phi_k(x). \qquad (5)$$

If Eq. 5 is multiplied on the left by $\phi_m(x)$ and integrated over the support of the basis functions, one obtains the matrix version of the Schrödinger equation in the variational approach:[15]

$$\boldsymbol{H}\boldsymbol{c} = E\boldsymbol{S}\boldsymbol{c}, \qquad (6)$$

where $H_{ij} = \langle \phi_j | \hat{H} | \phi_i \rangle$ and $S_{ij} = \langle \phi_j | \phi_i \rangle$. Eq. 6 is a *square* matrix equation with Hermitian matrices. Eq. 6 has real eigenvalues that are upper limits for exact energies[45] and converge to energy levels as the basis approaches completeness. This standard variational approach requires computing many integrals because elements of **H** and **S** are integrals. Integrals are also necessary to compute an energy from coupled cluster theory (where $E = \langle \Psi | \hat{H} | \Psi \rangle$ with Ψ the many body wavefunction).[21]

When using Eq. 6, it is common to compute some integrals (e.g. those for the KEO and *S*) exactly and others using quadrature; this means that quadrature approximations to integrals must be accurate. Using a mix of exact integrals and quadrature causes several problems:

(i) Basis functions must have amplitude wherever wavefunctions have significant amplitude. Quadrature points must be chosen so that quadrature approximations to integrals computed in the region in which wavefunctions have significant amplitude are accurate. This may require having quadrature points close to points at which a term in the Hamiltonian is singular. In the electronic problem, the Coulombic potential supports both core, atom-centered states, which require dense grids, as well as delocalized states which require extended quadrature grids.

(ii) To ensure that quadrature approximations to integrals are accurate, it is best to choose basis functions for which integrals of $\phi_k(x)V(x)\phi_{k\prime}(x)$ and $\phi_k(x)\phi_{k\prime}(x)$ are finite. This limits the space of possible functions. The basis is deemed complete if the $L^2$ norm $\int (\psi(x) - \sum_{k=1}^{N} c_k \phi_k(x))^2 dx \approx 0$.

(iii) If the Hamiltonian has singularities, it is tricky to find basis functions with which one can represent wavefunctions everywhere, which is necessary in a variational calculation. Singularities also make quadrature approximations to integrals inaccurate. For example, it is not practical to represent a nuclear cusp with plane waves. For vibrational calculations, the singularities (if any) are in the KEO and not the PES. For electronic calculations, the singularities are in the potential.

(iv) For some problems, the quadrature grid required to compute accurate potential and overlap matrix elements is larger than the basis size.[46,47] When solving the vibrational problem, if the quadrature grid is large enough, an analytic PES is necessary, as the number of points is too large to be computed *ab initio*.

Collocation is an alternative way to cast the Schrödinger equation into matrix form.[48-51] It also uses a basis expansion of the wavefunction, but the coefficients are determined by demanding that the Schrödinger equation be satisfied at a set of points $\{x_i\}, i = 1, \ldots, M$:

$$\sum_{k=1}^{N} c_k \hat{T} \phi_k(x_i) + V(x) \sum_{k=1}^{N} c_k \phi_k(x_i) = E \sum_{k=1}^{N} c_k \phi_k(x_i) \tag{7}$$

or in matrix form

$$\boldsymbol{Dc} + \boldsymbol{VFc} = E\boldsymbol{Fc}, \tag{8}$$

where $D_{ik} = \hat{T}\phi_k(x_i)$, $F_{ik} = \phi_k(x_i)$, $V_{ik} = \delta_{ik} V(x_i)$. The elements of the matrix $\boldsymbol{D}$ can be computed analytically[2-4,6-10,48,52-54] or numerically.[1,5,11] The numerical application of the KEO allows one to use the Cartesian KEO and apply it to *any* basis functions in *any* coordinates.[1,5,11] Equation 8 is a rectangular matrix equation because in general $M > N$. It can be approximately solved as a rectangular matrix pencil.[2,3,9,55] Alternatively, it can be converted to a square generalized eigenvalue problem, that can be solved exactly, by left multiplying by $\boldsymbol{F}^{\mathrm{T}}$ [1,3-5,11-13,56]

$$\boldsymbol{F}^T(\boldsymbol{D} + \boldsymbol{VF})\boldsymbol{c} = E\boldsymbol{F}^T\boldsymbol{Fc}. \tag{9}$$

The exact solutions of Eq. 9 are approximate solutions of Eq. 8. Several authors have noticed that when Eq. 9 is used, it is not necessary that matrix elements be good approximations for integrals.[57-59] This "rectangular collocation" approach has the following properties and advantages:

(i) No integrals are computed. Although $\boldsymbol{F}^T\boldsymbol{VF}$ has the same form as a quadrature approximation (with unit weights), **accurate energies can be computed without choosing points so that elements of $\boldsymbol{F}^T\boldsymbol{VF}$ are quadrature approximations to integrals.** For this reason, there is no need to use points in the entire region in which wavefunctions have significant amplitude.

(ii) Any basis functions can be used. There is no requirement that $\phi_k(x)V(x)\phi_{k'}(x)$ or $\phi_k(x)\phi_{k'}(x)$ be integrable. There is no requirement that basis functions have amplitude where wavefunctions have amplitude. We have recently shown that collocation works well even when the basis functions do not satisfy the integrability condition.[60]

(iii) Because energy levels can be computed with basis functions that have no amplitude at singularities (even when wavefunctions have significant amplitude at the singularities) and it is not necessary to approximate integrals, singularities cause no problems.[31]

(iv) The basis is deemed complete if, at the $M>N$ collocation points, $\psi(x_i) - \sum_{k=1}^{N} c_k \phi_k(x_i) \approx 0$, i.e. the series converges pointwise.

(v) Any point distribution can be used. Important parts of space can be emphasized. In fact, accurate energies can be computed using only points in a small region of space. For example, we have shown that levels are accurate even when the collocation set does not extend to the classical turning points or when the points are placed on low-dimensional slabs.[5,31]

(vi) As no integrals need to be converged, the required collocation point set is, in some cases, small enough that the potential can be directly evaluated at the points, obviating the need to fit a PES.[4,6-10]

(vii) When using Eq. 9, it is not necessary to store matrices of size $M \times N$, it is possible to store matrices only of dimension $N \times N$, see Ref. 1.

These are significant advantages which should be utilized in applications.

One disadvantage of collocation is the fact that the matrix $\boldsymbol{F}^T(\boldsymbol{D} + \boldsymbol{V}\boldsymbol{F})$ is not symmetric, which increases the cost of computing the eigenvalues. Eigenvalues that are not converged may be complex. Although collocation has the advantage that accurate spectra can be computed with fewer points than would be required with quadrature, when the point set is not dense, the wavefunction between the points might be poor, as the method only focuses on satisfying the Schrödinger equation at the points. Although many collocation point sets work well, for some points sets, the condition number of $\boldsymbol{F}^T\boldsymbol{F}$ is large and it is not possible to calculate accurate eigenvalues. Rectangular collocation is useful because there is a large middle ground between a point set with which quadrature approximations for elements of the overlap matrix are accurate and point sets for which the condition number of $\boldsymbol{F}^T\boldsymbol{F}$ is a problem. Rectangular collocation has been successfully applied to compute anharmonic molecular spectra of molecules with as many as 15 fully coupled degrees of freedom.[1-11] The relatively small size of the collocation point set and ability to compute the potential matrix without an analytic PES make it possible to compute anharmonic spectra for molecules on surfaces[6-8,10] which is useful in applications such as heterogeneous catalysis, photocatalysis, and electrochemical power sources.[61] Collocation has recently been used in multi-configuration time dependent Hartree[62] calculations.[63,64]

## 3  Methods

The Kohn-Sham Eq. 2 for CO and H$_2$O was solved using Eqs. 8, 9 with $\boldsymbol{x} = \{x, y, z\}$. The C-O bond length was set to 1.128 Å; the H-O bond length to 0.965 Å, and the HOH angle to 104 degrees. We solve the Kohn-Sham equation on a given effective potential computed with Gaussian 09.[65] Electrostatic potentials (ESP) and the density were output from single-point Gaussian 09 calculations as cube files with a resolution of 0.0945 Bohr and 200 points in each direction. The 6-311++g(2d,2p) basis was used. We used the Xα functional[66] to simplify constructing the exchange-correlation potential (which in the case of Xα is exchange-only):

$$V_x(\boldsymbol{x}) = 0.7 \frac{3}{2}\left(\frac{3}{\pi}\right)^{\frac{1}{3}} \rho(\boldsymbol{x})^{\frac{1}{3}}. \qquad (10)$$

Any exchange-correlation functional could be used; we chose Xα to avoid possible inaccuracies when extracting $V_{eff}$ from a DFT code and for direct comparison with the results of Ref. 31. The $V_x$ of Eq. (10) computed on the grid was added to the ESP to form $V_{eff}$.

Elements of the matrix $\boldsymbol{D}$ are computed with a five-point finite-difference (FD) stencil, with a FD step of $1 \times 10^{-6}$ *a.u*. The calculations were performed in Octave;[67] the generalized eigenvalue problem was solved using the *eig* function. The quality of the solution was evaluated vs. the reference levels and by monitoring the residual[31]

$$Res \equiv \frac{\langle |\psi| || H\psi - E\psi| \rangle}{\langle \psi | \psi \rangle} \approx \frac{|\boldsymbol{c}^T \boldsymbol{F}^T||(\boldsymbol{D}+(\boldsymbol{V}-E)\boldsymbol{F})\boldsymbol{c}|}{\boldsymbol{c}^T \boldsymbol{F}^T \boldsymbol{F} \boldsymbol{c}} \qquad (11)$$

for relevant levels. In Eq. 11, "| |" implies that the absolute values of the elements of vectors are used; it is not a norm. We use the residual of Eq. 11 because it is a rectangular residual vector, $(\boldsymbol{D} + (\boldsymbol{V} - E)\boldsymbol{F})\boldsymbol{c}$, weighted with the absolute value of the wavefunction. Energies with errors close to a mHa are achieved when *Res* is about 0.02 *a.u*.

The basis functions have the form $\phi_{k\zeta_{lm}}(\boldsymbol{x}) = \phi_\zeta(R_k) Y_{lm}(\boldsymbol{x} - \boldsymbol{x}_k)$, where $Y_{lm}$ is a spherical harmonic, $R_k = \|\boldsymbol{x} - \boldsymbol{x}_k\|$, and $\boldsymbol{x}_k$ is the 'center' of the $k^{\text{th}}$ set of basis function. The functions are centered on the ions. Multiple basis functions are positioned at each $\boldsymbol{x}_k$, and their number is controlled by a parameter $l_{max}$ such that $l = 0, 1, \ldots, l_{max}$, and $m = -l, -l+1, \ldots, l$.

Simple exponential functions $\phi_\zeta(R) = e^{-\epsilon_\zeta R}$ were used for radial components. The exponential form has the advantage that it has the correct form close to the nuclear cusp.[68] The parameter $\epsilon_\zeta$ determines the basis function width. At each center, we use several $\epsilon_\zeta$ which we denote, $\zeta = 1,…, \zeta_{max}$. That is, we use a multi-zeta type basis.[15] The values of $\epsilon_\zeta$ were chosen to approximately minimize *Res*.

We computed the lowest energy levels starting from core 1*s* levels of O and C and up to the molecular valence states (HOMO, highest occupied molecular orbital, and LUMO, lowest unoccupied molecular orbital). As identified in Ref. 31, to get millihartree accuracy for CO, seven $\zeta$-components were required as well as $l_{max} = 2$ (called an *spd* basis). The $\epsilon_\zeta$ values were 0.15, 0.195, 0.270, 0.315, 0.480, 0.810, and 1.350 (all in *a.u.*), multiplied by the value of Z, where Z is nuclear charge. For $H_2O$, six $\zeta$-components were required and $l_{max} = 2$ (an *spd* basis). The $\epsilon_\zeta$ values for O were 0.30, 0.39, 0.54, 0.63, 0.96, and 1.62 (all in *a.u.*), all multiplied by the value of Z, and 0.60, 0.78, 1.08, 1.26, 1.92, 3.24, for H. These parameters were manually selected to approximately minimize *Res* and were found to be relatively insensitive to collocation point distribution. The basis size was 126 for CO and 162 for $H_2O$.

Collocation points $x_i$ were selected from the uniform grid of 200×200×200 output by Gaussian 09. In general, collocation points can be distributed as one wishes, e.g. there is no need for them to be on a grid. They could be selected from any large pool of points, but the uniform grid is convenient because we are interfacing with *ab initio* calculations. First, point sets similar to those used in Ref. 31 were constructed by filtering all the points on the uniform grid. A grid point $x_i$ is accepted into a collocation point set if[1,5,11,69]

$$\frac{V_{max} - V(x_i) + \delta}{V_{max} - V_{min} + \kappa} \equiv \frac{V_{max} - V(x_i) + \delta}{V_{max} - V'} > rand, \qquad (12)$$

where *rand* is a random number in the range [0, 1]. Because there is a random element in Eq 12, the point set depends on the specific random sequence. This point selection scheme is similar to the simple potential-weighted scheme used for point selection in vibrational calculations[1,11] except that we introduce $\kappa$ to flatten the distribution between $V_{min}$ and $V' = V_{min} - \kappa$ and thereby ensure that the point set contains enough points between $V_{min}$ and $V'$. This flattening is helpful because the potential for a molecule has (see Figure 2) several deep minima separated by high barriers. Application of Eq. 12 with $V' = -35$ *a.u.*, $\delta = 0.2$ *a.u.* (for both molecules) resulted in a collocation set of about 51,000 points, which was large enough to achieve millihartree accuracy in Ref. 31. In this paper, the parameters $\kappa$ and $\delta$ are partially optimized. $\kappa$ is selected to ensure that all minima are sampled and that the point set obtained from Eq. 12 includes $M'$ = 1000, 2000, and 3000 points. $\delta$ is chosen by making plots and choosing $\delta$ values that minimize the mean absolute error (*mae*), on the full point set (about 51,000), of a potential represented with Gaussian Processes (GP) and obtained from a fitting set with $M'$ points. GP was programmed using Python's *scikit-learn* library.[70] The rational quadratic kernel was used. Near-optimal kernel parameters were determined by doing several trial regressions and kept fixed.

Once good values of $\kappa$ and $\delta$ have been identified, we use them and Eq. 12 to generate 48 sets with $M'$ = 1000 points, 48 sets with $M'$ = 2000 points and 48 sets with $M'$ = 3000 points. A genetic algorithm, home-coded in Python, is used to determine an optimal point set for each of these three populations. For each $M'$, the points undergo crossover and mutation for a number of generations. Crossover is done as follows: 1) select the best 24 datasets; 2) randomly select 24 sets of 2 parents from the best 24; 3) from each set of parents, make a new dataset whose points have a 50% probability of coming from each of the two parents. Each of these 24 new datasets then undergoes mutation by replacing some of its $M'$ points with points randomly selected from the uniform grid of 200×200×200. The "best" 24 are the sets for which the *rmse*, on the point set with about 51,000 points, of the GP potential is the smallest. Each of the $M'$ points has a mutation rate of 4% (probability to be mutated). Together, the best 24 sets and the 24 offspring generated from them constitute a new set of 48 points and a new "generation". The process of making generations is repeated until convergence is achieved. We find that evolution for about 30 generations is sufficient to optimize the point set. The set with the best fitness value is then used as the collocation point set to compute energy levels.

## 4 Results

### 4.1 *Optimization of parameters V' and $\delta$ of the probability density function*

Here we describe the procedure we use to choose a $\delta$ value for CO; a similar procedure is used for $H_2O$. We aim to achieve mHa accuracy with about 2,000 collocation points. We therefore select 2,000 points by applying Eq. 12 to all 8,000,000 cube points for different $V'$ and $\delta$ pairs. When a $V'$ and $\delta$ pair results in more than 2,000 points after application of Eq. 12, we select randomly 2,000 points from the resulting set. If fewer than 2,000 points are selected by Eq. 12, all those points are used. For each $V'$, $\delta$ pair and its set of $\leq$ 2,000 points, we compute the *rmse*, the *mae* (mean absolute error) of the GP effective potential at all the 51,000 points used in Ref. 31 as well as the residual *Res* of the KS equation for levels up to and including the LUMO. Figure 1 shows the dependence of the test-set *rmse, mae*, and of *Res* on *V'* and $\delta$. The *mae* and *Res* plots indicate that optimal values of $\delta$ around 0. We choose the values $\delta = 0$ and *V'* = -108 *a.u.*. See Figure 2 for a one-dimensional slice of the effective potential along the CO bond. The value of *V'* is chosen to be above -110 *a.u.* to make sure that the minimum of the effective potential well at the C atom (this minimum is finite due to discretization) is within the range of energies in which the probability of selection is unity, as well as to make sure that about 2,000 points are be obtained with Eq. 12 (lower / more negative values of $\kappa$ and $\delta$ result in fewer points). With $\delta = 0$ and *V'* = -108 *a.u.*, about 2,000 points are selected with Eq. 12. Similarly, the best 1,000 and 3,000 point sets were obtained by adjusting *V'* so that applying Eq. 12 to the 8,000,000 cube points gives, respectively, about 1,000 or 3,000 points.

The resulting point distribution, with $\delta = 0$ and *V'* = -108 *a.u.* and 2,000 points is shown in Figure 3. It is superimposed on the point distribution used in Ref. 31, with about 51,000 points. With the point set of Ref. 31, all the energy errors are about a mHa. They are about $10^{-2}$ Ha with the $\delta = 0$ and *V'* = -108 *a.u.* point set, although it contains many fewer points. The GA-optimized point set of the next subsection has a similar distribution, but better energy errors.

Table 1 shows the Gaussian 09 reference KS energies as well as energies and errors computed with collocation using point distributions obtained with Eq. 12 and different $\kappa$ and $\delta$ parameters. In column three of Table 1, we report energies computed with the $\kappa$ and $\delta$ parameters of Ref. 31, but using only 2,000 points. The errors are huge. Column 4 has energies computed with the "good" $\kappa$ and $\delta$ values, determined as explained in this subsection. Although the errors are *much* smaller, they are still unacceptably large. Even worse, the wavefunctions computed from such collocation point sets have extraneous features (see below). Much smaller errors (the GAGP columns) can be obtained, as presented in the next subsection, by using the probability density function determined by the best $\kappa$ and $\delta$ values as the starting point for an optimization with a GA.

Table 1. Kohn-Sham energies of CO computed with Gaussian 09 (reference) and with collocation on the same effective potential using different collocation point sets, constructed using the probability density function of Eq. 12 with particular $V'$, $\delta$ parameters without or with ("GPGA") subsequent application of machine learning (a combination of Gaussian process regression and a genetic algorithm). HOMO and LUMO energies are in bold. The sum of the eigenvalues of the occupied states and its error as well as the *rmse* of the GP-interpolated $V_{eff}$ are given at the bottom. The errors in the KS levels and in the sum vs. the reference are also given ("error"). All values are in *a.u*. The values in the columns labelled by "2,000 (Eq 12)" have poor reproducibility and should be viewed as representative.

| no. of points | 51,000 (Eq. 12) | | 2,000 (Eq. 12) | | 2,000 (Eq. 12) | | 1,000 (GAGP) | | | | 2,000 (GAGP) | | | | 3,000 (GAGP) | | | |
|---|---|---|---|---|---|---|---|---|---|---|---|---|---|---|---|---|---|---|
| | | | | | | | Generation 29 | | Final generation | | Generation 22 | | Final generation | | Generation 21 | | Final generation | |
| | $\kappa$ | $\delta$ | $\kappa$ | $\delta$ | $\kappa$ | $\delta$ | $\kappa$ | $\delta$ | $\kappa$ | $\delta$ | $\kappa$ | $\delta$ | $\kappa$ | $\delta$ | $\kappa$ | $\delta$ | $\kappa$ | $\delta$ |
| | -35.00 | 0.20 | -35.00 | 0.20 | -108.00 | 0.00 | -208.00 | 0.00 | -208.00 | 0.00 | -108.00 | 0.00 | -108.00 | 0.00 | -70.00 | 0.00 | -70.00 | 0.00 |
| G09 energies | energy | error | energy | error | energy | error | energy | error | energy | error | energy | error | energy | error | energy | error | energy | error |
| -18.7401 | -18.7384 | 0.0017 | -36.3609 | -17.6208 | -18.7270 | 0.0131 | -18.7550 | -0.0146 | -18.8100 | -0.0696 | -18.7430 | -0.0032 | -18.7400 | -0.0003 | -18.7390 | 0.0009 | -18.7350 | 0.0051 |
| -9.9072 | -9.9037 | 0.0035 | -24.2358 | -14.3286 | -9.8568 | 0.0504 | -9.9171 | -0.0099 | -9.9156 | -0.0084 | -9.9075 | -0.0003 | -9.9099 | -0.0027 | -9.9041 | 0.0031 | -9.9082 | -0.0010 |
| -1.0480 | -1.0410 | 0.0070 | -23.6071 | -22.5590 | -1.0491 | -0.0011 | -1.0522 | -0.0042 | -1.0624 | -0.0144 | -1.0467 | 0.0014 | -1.0503 | -0.0023 | -1.0485 | -0.0004 | -1.0478 | 0.0002 |
| -0.4912 | -0.4922 | -0.0010 | -23.6071 | -23.1158 | -0.4917 | -0.0005 | -0.4915 | -0.0002 | -0.4926 | -0.0014 | -0.4903 | 0.0009 | -0.4905 | 0.0008 | -0.4920 | -0.0007 | -0.4909 | 0.0004 |
| -0.4133 | -0.4162 | -0.0029 | -20.4418 | -20.0286 | -0.4163 | -0.0030 | -0.4167 | -0.0034 | -0.4279 | -0.0146 | -0.4149 | -0.0017 | -0.4190 | -0.0057 | -0.4157 | -0.0024 | -0.4138 | -0.0005 |
| -0.4133 | -0.4142 | -0.0009 | -20.4418 | -20.0286 | -0.4143 | -0.0010 | -0.4144 | -0.0011 | -0.4100 | 0.0033 | -0.4116 | 0.0017 | -0.4151 | -0.0018 | -0.4129 | 0.0004 | -0.4121 | 0.0012 |
| **-0.3043** | **-0.3045** | **-0.0002** | **-17.8687** | **-17.5644** | **-0.3049** | **-0.0006** | **-0.3056** | **-0.0013** | **-0.3055** | **-0.0012** | **-0.3044** | **-0.0001** | **-0.3058** | **-0.0015** | **-0.3046** | **-0.0003** | **-0.3054** | **-0.0012** |
| **-0.0518** | **-0.0529** | **-0.0011** | **-17.8687** | **-17.8169** | **-0.0520** | **-0.0001** | **-0.0527** | **-0.0009** | **-0.0540** | **-0.0022** | **-0.0526** | **-0.0008** | **-0.0541** | **-0.0022** | **-0.0530** | **-0.0011** | **-0.0523** | **-0.0005** |
| -0.0518 | -0.0519 | -0.0001 | -17.7264 | -17.6745 | -0.0517 | 0.0001 | -0.0509 | 0.0009 | -0.0488 | 0.0030 | -0.0513 | 0.0005 | -0.0528 | -0.0010 | -0.0524 | -0.0005 | -0.0517 | 0.0001 |
| 0.0384 | 0.0619 | 0.0235 | -17.7264 | -17.7648 | 0.0893 | 0.0509 | 0.0864 | 0.0480 | 0.0976 | 0.0592 | 0.0822 | 0.0438 | 0.0827 | 0.0443 | 0.0931 | 0.0547 | 0.0901 | 0.0517 |
| -31.3174 | -31.3102 | 0.0072 | -166.5632 | -135.2458 | -31.2601 | 0.0573 | -31.3525 | -0.0351 | -31.4239 | -0.1065 | -31.3185 | -0.0011 | -31.3305 | -0.0131 | -31.3168 | 0.0006 | -31.3132 | 0.0042 |
| | | | | | | | *rmse* of $V_{eff}$ | | | | | | | | | | | |
| | | | | | 0.0019 | | 0.0019 | | 0.0009 | | 0.0007 | | 0.0018 | | 0.0013 | | | |

Table 2. Kohn-Sham energies of $H_2O$ computed with Gaussian 09 (reference) and with collocation on the same effective potential using different collocation point sets, constructed using the probability density function of Eq. 12 with particular $V'$, $\delta$ parameters without or with ("GPGA") subsequent application of machine learning (a combination of Gaussian process regression and a genetic algorithm). HOMO and LUMO energies are in bold. The sum of the eigenvalues of the occupied states and its error as well as the *rmse* of the GP-interpolated $V_{eff}$ are given at the bottom. The errors in the KS levels and in the sum vs. the reference are also given ("error"). All values are in *a.u.* The values in columns labelled by "2,000 (Eq. 12)" have poor reproducibility and should be viewed as representative.

| no. of points | 51,000 (Eq. 12) | | 2,000 (Eq. 12) | | 2,000 (Eq. 12) | | 1,000 (GAGP) | | | | 2,000 (GAGP) | | | | 3,000 (GAGP) | | | |
|---|---|---|---|---|---|---|---|---|---|---|---|---|---|---|---|---|---|---|
| | | | | | | | Generation 12 | | Final generation | | Generation 18 | | Final generation | | Generation 5 | | Final generation | |
| | $\kappa$ | $\delta$ | $\kappa$ | $\delta$ | $\kappa$ | $\delta$ | $\kappa$ | $\delta$ | $\kappa$ | $\delta$ | $\kappa$ | $\delta$ | $\kappa$ | $\delta$ | $\kappa$ | $\delta$ | $\kappa$ | $\delta$ |
| | -35.00 | 0.20 | -35.00 | 0.20 | -108.00 | 0.00 | -208.00 | 0.00 | -208.00 | 0.00 | -108.00 | 0.00 | -108.00 | 0.00 | -70.00 | 0.00 | -70.00 | 0.00 |
| G09 energies | energy | error | energy | error | energy | error | energy | error | energy | error | energy | error | energy | error | energy | error | energy | error |
| -18.6306 | -18.6329 | -0.0023 | -61.6910 | -43.0600 | -18.6480 | -0.0171 | -18.6360 | -0.0049 | -18.6470 | -0.0164 | -18.6350 | -0.0039 | -18.6580 | -0.0274 | -18.6390 | -0.0087 | -18.6480 | -0.0171 |
| -0.8931 | -0.8952 | -0.0021 | -55.4390 | -54.5460 | -0.8987 | -0.0056 | -0.8893 | 0.0039 | -0.8784 | 0.0147 | -0.8987 | -0.0055 | -0.8941 | -0.0010 | -0.8953 | -0.0022 | -0.8941 | -0.0009 |
| -0.4521 | -0.4524 | -0.0003 | -40.7840 | -40.3310 | -0.4646 | -0.0125 | -0.4442 | 0.0079 | -0.5170 | -0.0649 | -0.4486 | 0.0035 | -0.4570 | -0.0049 | -0.4514 | 0.0007 | -0.4613 | -0.0092 |
| -0.3140 | -0.3210 | -0.0070 | -29.1990 | -28.8850 | -0.3197 | -0.0057 | -0.2976 | 0.0165 | -0.2770 | 0.0371 | -0.3201 | -0.0060 | -0.3132 | 0.0009 | -0.3169 | -0.0028 | -0.3197 | -0.0057 |
| **-0.2378** | **-0.2305** | **0.0073** | **-25.3060** | **-25.0680** | **-0.2175** | **0.0203** | **-0.2227** | **0.0151** | **-0.2109** | **0.0269** | **-0.2353** | **0.0026** | **-0.2224** | **0.0154** | **-0.2384** | **-0.0006** | **-0.2263** | **0.0116** |
| **-0.0138** | **-0.0170** | **-0.0032** | **-25.3060** | **-25.2920** | **-0.0142** | **-0.0004** | **-0.0155** | **-0.0018** | **-0.0198** | **-0.0061** | **-0.0156** | **-0.0018** | **-0.0164** | **-0.0026** | **-0.0150** | **-0.0012** | **-0.0148** | **-0.0010** |
| 0.0391 | 0.0194 | -0.0197 | -24.5430 | -24.5820 | 0.0404 | 0.0013 | 0.0355 | -0.0035 | 0.0321 | -0.0070 | 0.0377 | -0.0014 | 0.0418 | 0.0028 | 0.0403 | 0.0012 | 0.0397 | 0.0006 |
| 0.1407 | 0.0239 | -0.1168 | -24.5430 | -24.6840 | 0.0639 | -0.0767 | 0.0552 | -0.0854 | 0.0594 | -0.0812 | 0.0574 | -0.0833 | 0.0594 | -0.0813 | 0.0614 | -0.0792 | 0.0616 | -0.0791 |
| -20.5276 | -20.5320 | -0.0044 | -212.4190 | -191.8914 | -20.5485 | -0.0209 | -20.4897 | 0.0379 | -20.5303 | -0.0027 | -20.5375 | -0.0099 | -20.5446 | -0.0170 | -20.5409 | -0.0133 | -20.5493 | -0.0217 |
| | | | | | | | *rmse* of $V_{eff}$ | | | | | | | | | | | |
| | | | | | | | 0.0036 | | 0.0026 | | 0.0028 | | 0.0025 | | 0.0043 | | 0.0016 | |

*4.2 Machine learning optimization of point distribution*

CO energies computed with the collocation point set refined with the GA using the accuracy of the GP approximation to the potential are also shown in Table 1. Using only 2,000 points, errors of about one mHa are obtained. For comparison, energies with 1,000 and 3,000 collocation points are also given. They were computed from the GP+ GA optimization, starting with points obtained from Eq. 12 with $\kappa$ and $\delta$ values determined as explained in the previous section. With only 1,000 points, it was not possible to achieve mHa accuracy; with 3,000 points errors are about $10^{-3}$ *a.u.* for both the levels and the sum of eigenvalues for the occupied levels (the bandstructure part of the total energy). We observe that the accuracy of the energies is not a monotonic function of the accuracy of the potential. For example, better energies are obtained from intermediate GA generations (see e.g. generation 22 in Table 1 in the 2,000 point case) than in the final generation that has a slightly smaller potential error. This highlights the limits of the strategy of minimizing potential error when the density of the data is low. Nevertheless, mHa accuracy of the CO KS spectrum is maintained through final generations.

We used a similar strategy to optimize the collocation point set for $H_2O$. Similar values of $\kappa / V'$ and $\delta$ were needed to select 1,000 / 2,000 / 3,000 points from the total set of 8,000,000, see Table 2. For $H_2O$, we do not choose $\kappa$ so that all points close to a H nucleus are accepted by Eq. 12 because H $1s$ levels do not preserve their character in the molecule, contrary to $1s$ levels of C and O atoms in CO. Similar to the CO case, with 2,000 points, energy errors are about $10^{-2}$ Ha, before the point set is refined with the GA. After the GA optimization, and with the same number of points, the errors are about one mHa. Similar to the CO case, we observe that intermediate generations give slightly better energies, even though in the final generation the *rmse* of the effective potential is slightly lower. The sum of occupied eigenvalues is accurate to $10^{-3}$ Ha in the best generations. Contrary to the case of CO, the mHa accuracy is not maintained in the final generations, and this issue persists when increasing the number of points to 3,000. The limitations of the strategy of minimizing the *rmse* of the potential to improve the levels appear to be more severe in the case of $H_2O$. In the future, alternative optimization criteria could be explored.

Several molecular orbitals of CO and $H_2O$, starting with the lowest-energy orbitals and through the LUMO (LUMO+1 in the case of CO as LUMO and LUMO+1 are degenerate) computed by solving the Kohn-Sham equation using 2,000 GA-GP optimized collocation points are shown in Figure 4 and Figure 5, respectively. They are visually indistinguishable from those computed by Gaussian and by the collocation method using the large set of 51,000 points, see Fig. 6 of Ref. 31. The data for Figure 4 and Figure 5 were computed with Eq. 4 on the same $200^3$ cube points originally generated with Gaussian. Although there is no guarantee that wavefunctions computed with collocation will be accurate at points that are not collocation points, with a sufficiently dense collocation point set this is the case, as for example with the 51,000 point set used in Ref. 31. It is also the case with the small GA-optimized point sets, but not necessarily the case with the non-optimized sets. For example in Figure 6, the orbital plots of $H_2O$ computed from a set of 2,000 points with the same $\delta = 0$ and $V' = -108$ *a.u.* but without machine learning optimization, have easily visible deficiencies. The corresponding residuals are large (on the order of 3 *a.u.*) although the energies have errors only on the order of $10^{-2}$ *a.u.* (Table 2).

## 5 Conclusions

We have used a genetic algorithm to optimize the collocation point set used to solve the electronic Schrödinger equation with the rectangular collocation approach. Other ideas have been used before to identify the most important region of a potential surface when studying nuclear dynamics. Here, we used machine learning for point optimization.[71,72] The optimization criterion is the accuracy of a Gaussian Processes regression potential made from the points. Using an optimized set of 2,000 points to solve the Kohn-Sham equations for CO and $H_2O$ with a reference effective potential yields orbital energies and a sum of energies with errors of about a mHa. This point set is more than an order of magnitude smaller than the point set employed in Ref. 31, indicating that points that are good for representing the potential are also good for computing energies. This is useful because it means that the potential can be used to refine the point set without solving the Schrödinger equation. We fed the GA populations generated by choosing points using a probability distribution with parameters optimized with the GP interpolated potential and large set of test points. We observe that although it is helpful to start with a good probability density function, optimizing its parameters is not by itself sufficient to achieve mHa accuracy: it is necessary to also do the GA refinement.

A key advantage of rectangular collocation is the possibility to compute accurate energies using small basis sets and basis functions that have no amplitude at singularities and do not satisfy integrability conditions. Another advantage is that one may use a small collocation point set and that many point distributions work well. All of these advantages stem from the fact that no integrals are computed. In this paper, we show that machine learning can be used to significantly reduce the size of the collocation point set, without loss of accuracy. It is possible that by using a small basis and an optimized point set with rectangular collocation it will be less costly to solve the Schrödinger equation than with standard methods.

The idea of choosing points for solving the Schrödinger equation by optimizing the accuracy of an interpolant for the potential is interesting. For the molecules studied in this paper, it works, but is not perfect. We observe that slightly *better* energies are obtained with point sets having some slightly *larger* potential *rmse* values. For CO, mHa accuracy was then maintained through later generations of the GA optimization, but the accuracy did not improve as the potential *rmse* slightly improved. For the water molecule, mHa accuracy was achieved in intermediate generations, but lost in later generations, with a couple levels having errors as large as $10^{-2}$ *a.u.,* although the potential *rmse* did slightly improve. This is indicative of the limits of using potential errors to guide the choice of collocation points when the points are not dense enough and highlights the need to explore other optimization criteria. Overall, however, potential-driven GA optimization is effective. For both CO and $H_2O$, GA-optimized collocation point sets resulted in correct orbital shapes and accurate energies and with no GA optimization, although energies are relatively accurate, points drawn from the same distribution can yield bad orbitals (incorrect behavior between collocation points) and high rectangular residuals.

# 6 Acknowledgements

This work was supported by the Natural Science and Engineering Research Council of Canada. A. K. thanks Strategic India Initiative by the National University of Singapore.

# 7 Conflicts of interest

There are no conflicts of interest to declare.

## 9  Figures

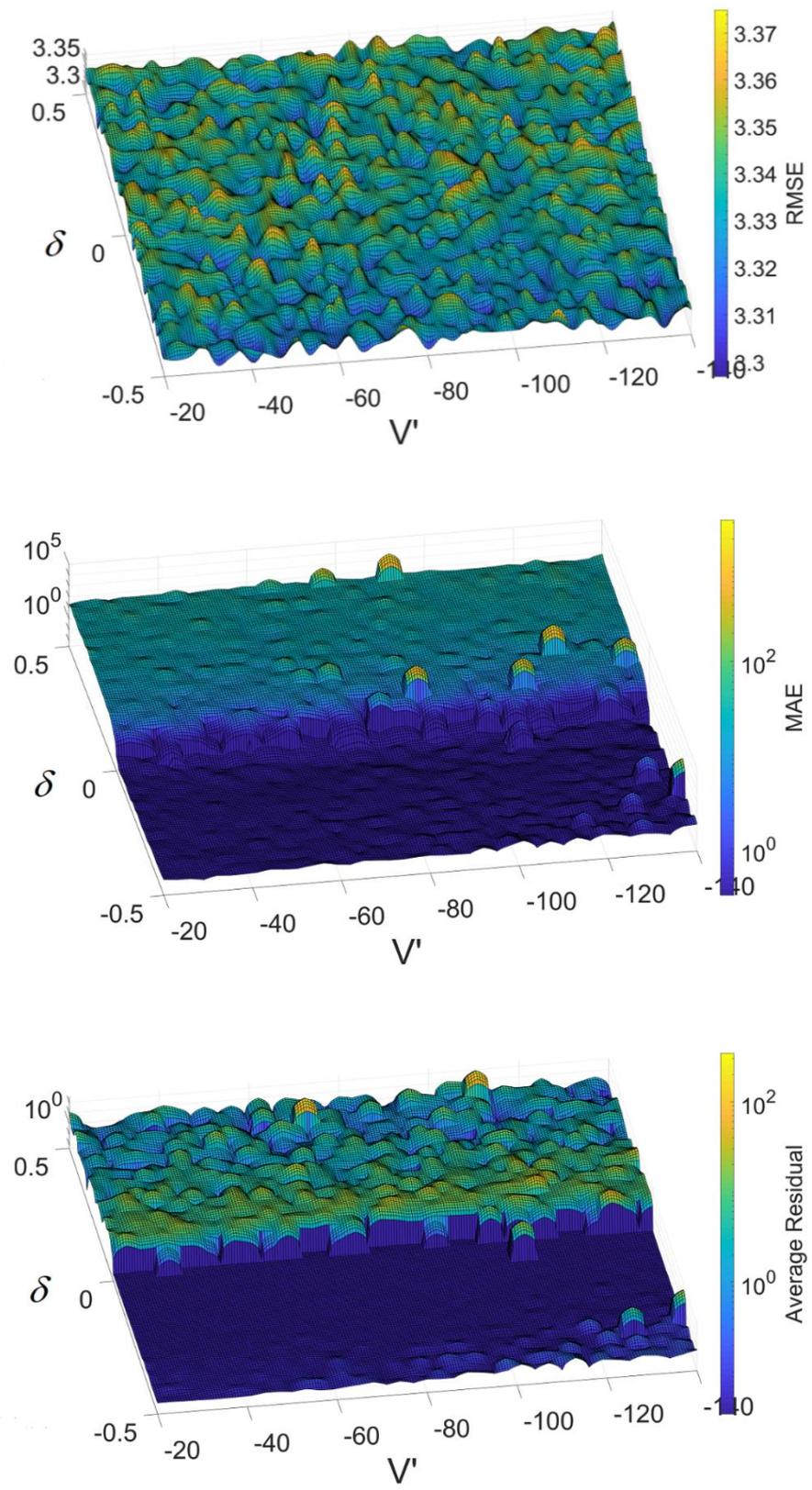

Figure 1 Top to bottom: the interpolated potential test set *rmse* (top panel), *mae* (middle panel), and the collocation equation residual (Eq. 11, bottom panel) for CO when randomly selecting at most 2,000 points with Eq. 12.

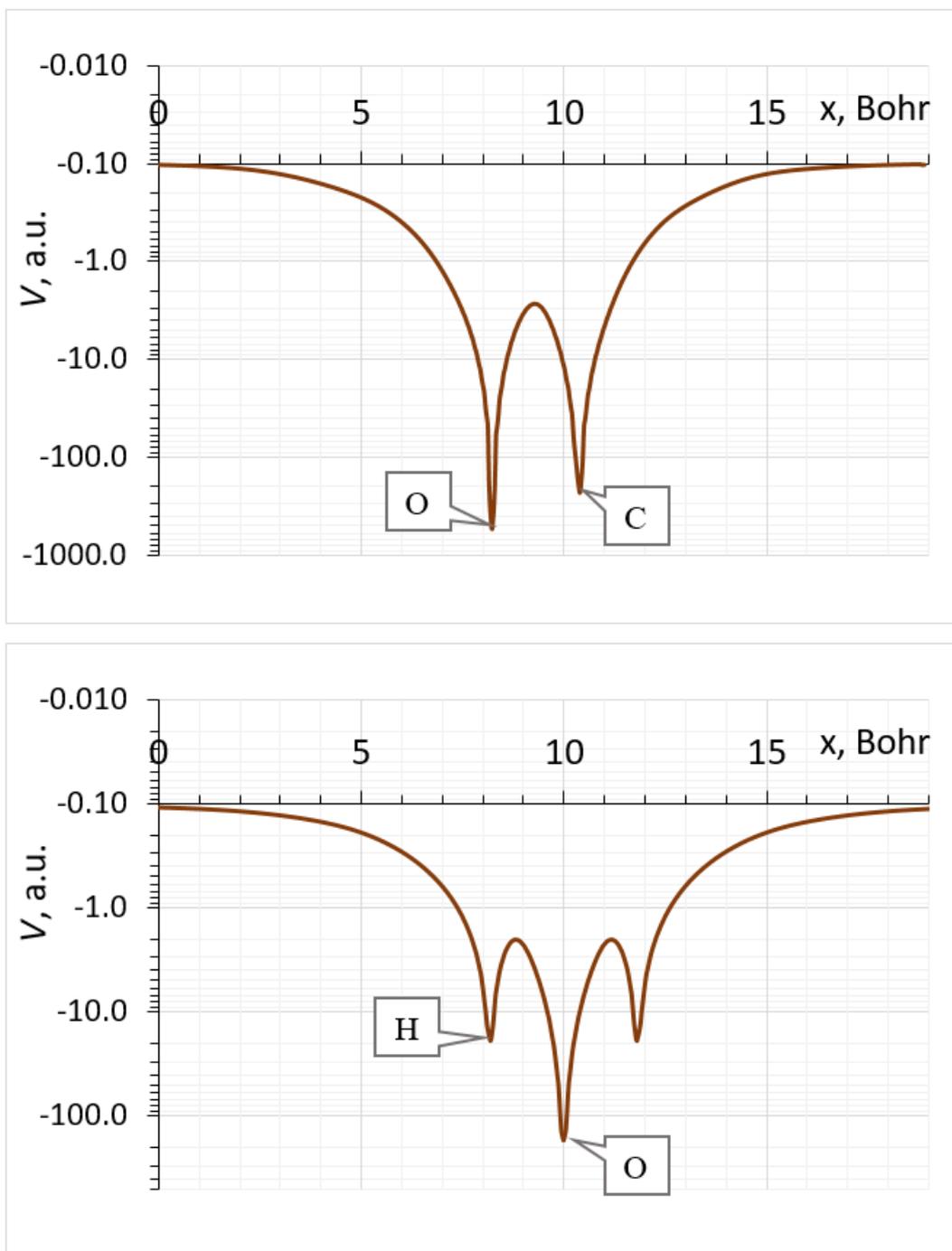

Figure 2 A scan of CO (top panel) and $H_2O$ (bottom panel) effective potential along the CO and OH bonds, respectively. An offset of 0.1 *a.u.* is applied to plot on the logarithmic scale. Finite well depths are due to the finite cube resolution.

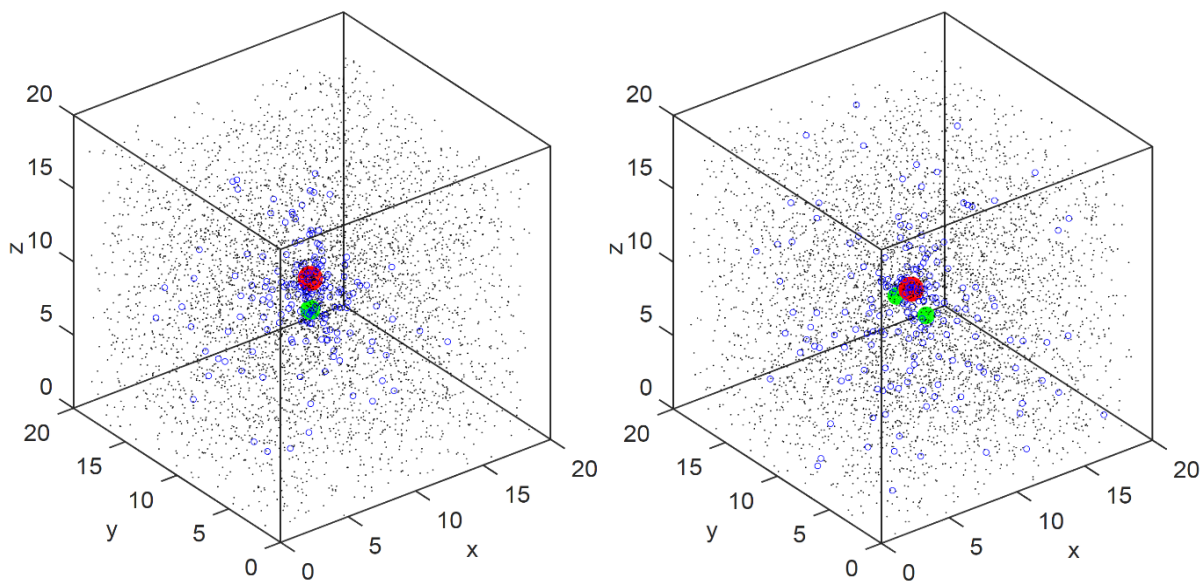

Figure 3 Distributions of collocation points for CO (left panel) and $H_2O$ (right panel): black dots – the 51,000 point distribution used in Ref. 31, blue circles – the 2,000 point distribution used here. Random subsets of points (rather than all 51,000 and 2,000 points) are shown for better visibility. C and O atoms are shown as green and red spheres, respectively, for CO. H and O atoms are shown as green and red spheres, respectively, for $H_2O$.

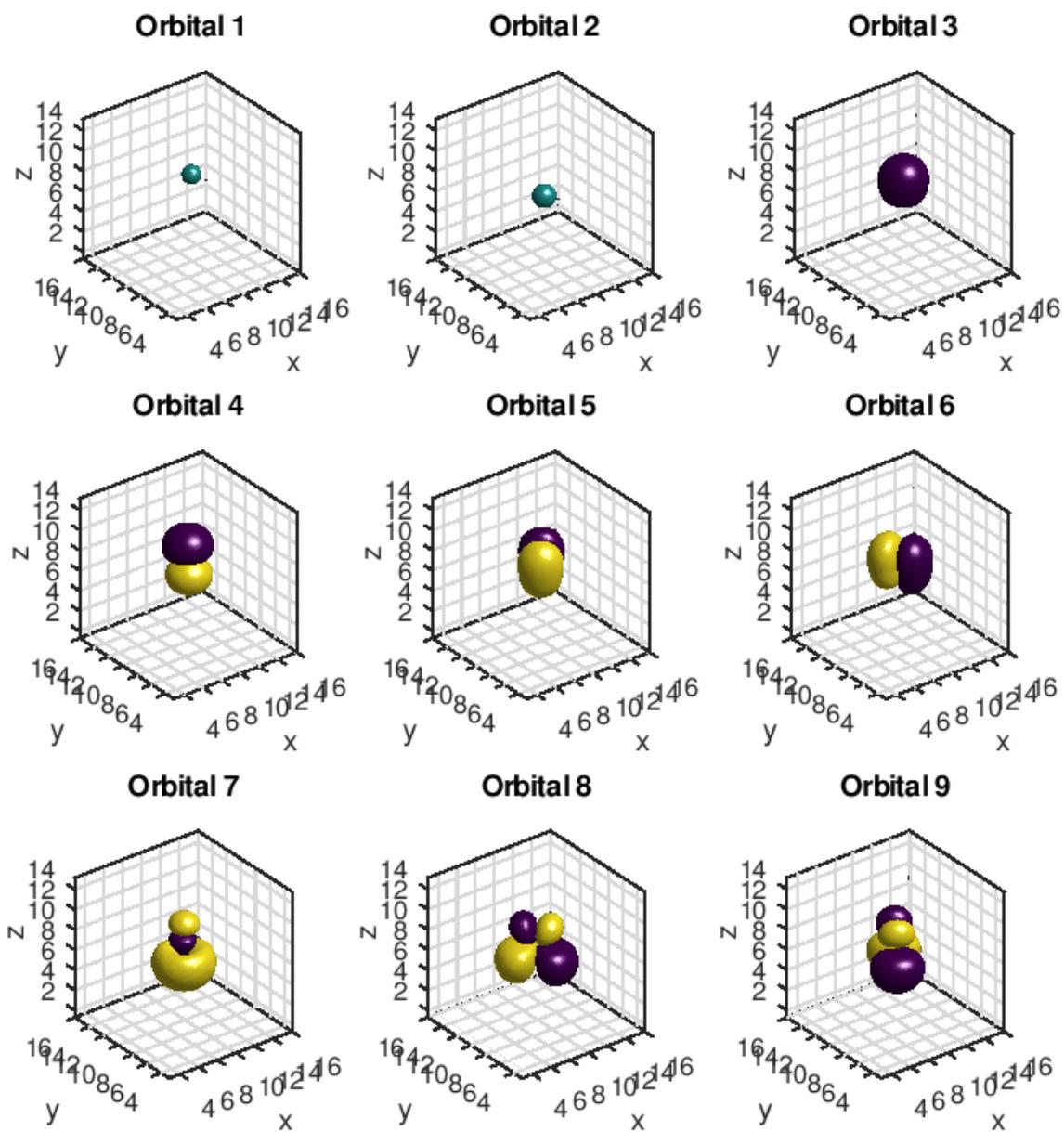

Figure 4 The lowest-energy orbitals of the CO molecule plotted from the solution of the Kohn-Sham equation using 2,000 GA-GP optimized collocation points. Orbitals 7 and 8 are HOMO and LUMO, respectively.

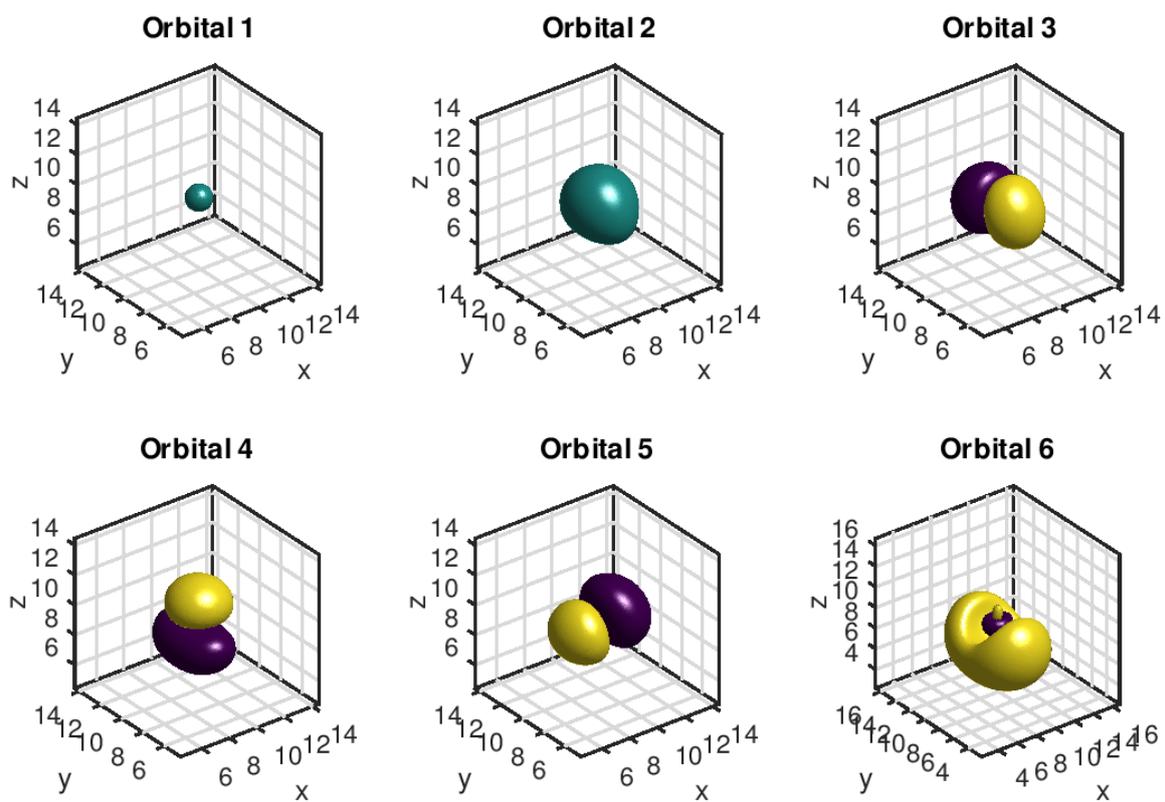

Figure 5 The lowest-energy orbitals of the $H_2O$ molecule plotted from the solution of the Kohn-Sham equation using 2,000 GA-GP optimized collocation points. Orbitals 5 and 6 are HOMO and LUMO, respectively.

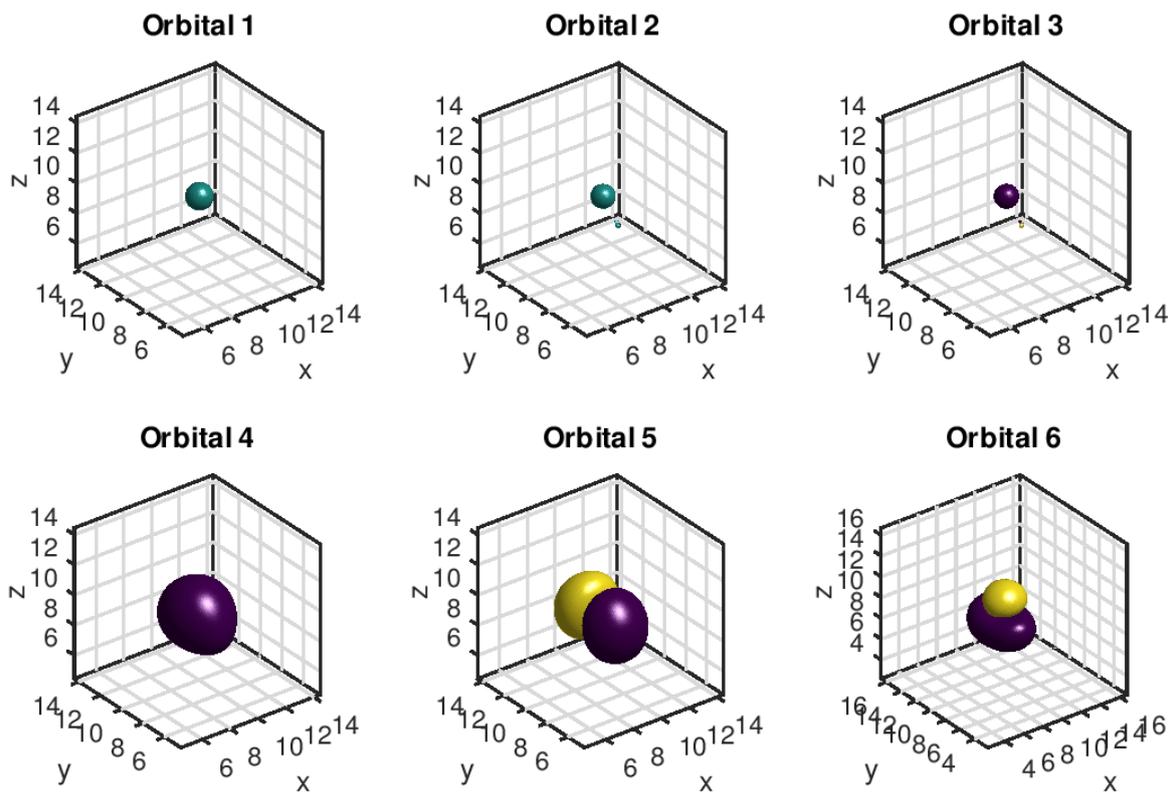

Figure 6 The lowest-energy orbitals of H$_2$O plotted from solutions of the Kohn-Sham equation using 2,000 *non*-optimized collocation points.